\documentstyle[12pt]{article}
\begin{document}
\begin{titlepage}
\begin{flushright}
UAB--FT--402\\
L.P.T.H.E.--ORSAY 96/88\\
October 1996\\
hep-ph/9610205
\end{flushright}
\vspace{1cm}

\begin{center}
{\bf\LARGE Leptonic Photons and Nucleosynthesis}
\vspace{2cm}

\centerline{\bf J. A. Grifols$^{a}$ and
                E. Mass\'o$^{a,b}$}
\vspace{0.6cm}

\centerline{$^a$Grup de F\'{\i}sica Te\`orica and IFAE,
                Universitat Aut\`onoma de Barcelona,}
\centerline{E-08193 Bellaterra, Spain}
\centerline{$^b$LPTHE, b\^at. 211, Universit\'e Paris XI, Orsay Cedex, 
France$^{\ast}$}
\vspace{3cm}

{\bf Abstract}\\
\parbox[t]{\textwidth}
{Should $U(1)$ long-range forces be associated to electron, muon
and/or tau quantum number then their ''fine structure constants" are
seen to be bound by nucleosynthesis data to be less than about 
$1.7 \times 10^{-11}$. For $\tau$ and $\mu$ this is the best upper  
limit up to date.}
\end{center}
\vspace{\fill}

{\noindent\makebox[10cm]{\hrulefill}\\
\footnotesize
\makebox[1cm][r]{$^{\ast}$}Laboratoire associ\'e au C.N.R.S. --- URA D0063.
}
\end{titlepage}
\newpage
 
While the r\^ole of electric charge, weak isospin, and color is completely 
clarified in the modern Particle Physics paradigm, i.e. they are sources
of interactions linked to local gauge invariance, this is not the case
for lepton and/or baryon number. Apparently, lepton (and baryon) number
conservation holds to an extremely high degree and no violation has
been experimentally established so far. Yet no forces associated
to these quantum numbers have been identified, a fact that clearly
distinguishes them from, say, electric charge. The question whether
baryon/lepton number are charges like electric charge or color (i.e.
which can be associated to forces that feel such charges) has already a
long tradition in Particle Physics. It was first raised by Lee and Yang
\cite{ly} who pointed out that the existence of these forces would spoil
the equality of gravitational and inertial mass. Very recently, Okun 
\cite{okun} has reexamined the whole issue of long-range forces associated
to electron, muon, and tau lepton number. He considers the existence of new
massless U(1) vector bosons coupled to those charges, considered as
independently conserved quantities. He analyses the various
phenomenological consequences of their presence and states the
corresponding limits on the associated ''fine structure constants",
$\alpha_e$, $\alpha_\mu$, and $\alpha_\tau$.

The properties of the tau lepton doublet are in general the less precisely
known among leptons and this turns out to be also the case for the issue
discussed here. Indeed, as Okun states in his paper \cite{okun}, there are
no direct laboratory upper limits on $\alpha_\tau$ and the limit on
$\alpha_\mu$ ($\alpha_\mu  < 10^{-5}$) is much worse than $\alpha_e <
10^{-49}$ for the first generation, derived from Dicke's experimental
limit on the equivalence between inertia and gravity. In the present paper
we fill this void and provide a bound on $\alpha_\tau$ and $\alpha_\mu$. We
find $\alpha_{\tau, \mu} < 1.7 \times 10^{-11}$. These limits are obtained
from the nucleosynthesis constraint on extra effective massless degrees of
freedom and are much better than the laboratory limits stated above,
especially for the tau lepton. Since we consider electron, muon, and tau
lepton number as independent charges, our results are valid for either one
of them. For ease of presentation, however, we shall in what follows only
refer to the tau family.

Results on constraints derived from nucleosynthesis are frequently
presented in terms of how many equivalent massless neutrino species do data
on helium-4 and other light element abundances actually allow. A very
recent reanalysis of this question fixes this maximum number to be very
safely four \cite{four}. Therefore, at most one extra equivalent neutrino
species can be accomodated. Adding one $\tau$-photon ($\gamma_\tau$), and
the right-handed $\nu_\tau$ -also in equilibrium because of the vector
character of the hypothesized new interaction- would clearly exceed this
limit. Thus, $\tau$-photons should be decoupled for $T \sim O(1)$ MeV, the
neutrino decoupling temperature. 

Let us see the implication for the problem at hand. Decoupling of a
species occurs whenever the Hubble expansion rate overcomes the
annihilation rate for this species.  The cross-section for 
$\gamma_\tau\gamma_\tau \rightarrow \nu_\tau \nu_\tau$, in the C.M.,
is              
\begin{equation}
\label{sigma}
\sigma = {2 \pi \alpha_\tau^2  \over s }\, \left( \log{s \over k_D^2 }
- 1 \right)   \ , 
\end{equation}
where $s$ is the C.M. energy squared and $k_D$ is the Debye momentum, i.e.
the cut-off in momentum corresponding to the physical screening of the
interaction range by the $\tau$-charges of the neutrino-antineutrino
plasma. It is explicitly given by,  \begin{equation} \label{debye} 
k_D^2 = {\alpha_\tau n(\nu) \over T } 
\end{equation}
with $n(\nu)$ the number density of neutrinos. 
 
The relevant interaction rate is          
\begin{equation}
\label{gamma1}
\Gamma = n(\gamma_\tau) < \sigma v > 
\end{equation}
where the cross-section is thermalised and $n(\gamma_\tau)$ stands for the
$\gamma_\tau$ number density. A fairly good numerical approximation for it
is
\begin{equation}
\label{gamma2}
\Gamma = 0.053\, \alpha_\tau^2\, T\, \log{160 \over \alpha_\tau }
\end{equation}

On the other hand, the Hubble expansion rate is 
\begin{equation}
\label{H}
H = \sqrt{4 \pi^3 \over 45}\,  g_*^{1/2}\, {T^2 \over M_{Pl} } 
\end{equation}
with $g_*(T)$ the number of effective degrees of freedom at temperature
$T$.

As a consequence of the $T$ dependence in Eqs.\,(\ref{gamma2}) and
(\ref{H}), if at a certain $T$ $\tau$-photons are in equilibrium
with neutrinos, they will remain in equilibrium for any temperature below
$T$ \footnote{Note that this is opposite to what happens with ordinary
weak interactions.}. Therefore, since for $T \sim O(1)$ MeV
$\tau$-photons  should be already frozen out, they were never before in
equilibrium with the ordinary relativistic plasma. We are forced to
require that equilibrium sets in for   $T < 1$ MeV. This requirement then
leads to   
\begin{equation}
\label{alpha}
\alpha_\tau  \leq 1.7 \times 10^{-11} 
 \end{equation}
by equaling Eqs.\,(\ref{gamma2}) and (\ref{H}) at $T = 1$ MeV. A larger 
$\alpha_\tau$ would imply equilibrium at  $T > 1$ MeV which is forbidden by
observation. Below this temperature, $\tau$-photons and neutrinos share the
same temperature, their wavelengths redshifting with the expansion so
that, at present, $\tau$-photons would contribute a fraction of the total
neutrino energy density.

\section*{Acknowledgments}

We acknowledge financial support from the CICYT AEN95-0815 and AEN95-0882
Research Projects and from the Theoretical Astroparticle Network under the
EEC Contract  No. CHRX-CT93-0120 (Direction Generale 12 COMA).
E.~M.~ack\-nowledges financial support from the Direcci\'on General de 
Investigaci\'on Cient\'{\i}fica y Ense\~nanza Superior (DGICYES) and from 
Comissionat per Universitats i Recerca de la Generalitat de Catalunya.
E.~M.~is grateful to Michel Fontannaz, Lluis Oliver, Olivier P\`ene, and 
to the Laboratoire de Physique Th\'eorique et Hautes Energies of the 
Universit\'e Paris XI for their hospitality while part of this work was 
performed.

\end{document}